\documentclass[dvips]{article}
\usepackage[dvips]{graphicx}
\title{
Low-Energy Predictions of Lopsided Family Charges
}

\author{
 J.~Sato\\%
{\footnotesize \it%
Research Center for Higher Education, Kyushu University,
Ropponmatsu, Chuo-ku, Fukuoka, 810-8560, Japan%
}
\\
and
\\
T.~Yanagida\\%
{\footnotesize \it%
 Department of Physics and RESCEU,
 University of Tokyo, Hongo, Bunkyo-ku, Tokyo
 113-0033, Japan%
}
}

\begin{document}

\maketitle
\abstract{%
We consider the Froggatt-Nielsen (FN) mechanism reproducing the observed 
mass hierarchies and mixing angles for quarks and leptons. The large 
$\nu_\mu $-$ \nu_\tau$ mixing suggested from recent Superkamiokande
experiments on the atmospheric neutrinos implies lopsided FN U(1)
charges for the lepton doublets. There are two possible charge
assignments to generate the large $\nu_\mu$-$ \nu_\tau$ mixing. We point
out that the two models with different charge assignments have distinct
low-energy predictions and hence they are distinguishable in future
neutrino experiments on such as CP violation and $2\beta 0\nu$ decay.
}

\section{Introduction}
The observed quark-lepton mass spectra and mixing angles may provide us
with important information on a more fundamental theory beyond the
standard model. There have been proposed many mechanisms and symmetries
to account for the observed mass spectra and mixings for quarks and
leptons. Among them we consider that the Froggatt-Nielsen(FN)
mechanism\cite{FN} using a broken U(1) family symmetry is the simplest
and the most promising candidate. In this scheme the masses and mixings
for quarks are well understood by choosing properly the FN U(1) charges
for each quarks.

The large $\nu_\mu$ -$\nu_\tau$ mixing observed by Superkamiokande
experiments on atmospheric neutrinos\cite{SK} has led us to propose
lopsided U(1) charges to the left-handed lepton doublets $l_i
(i=1-3)$\cite{SY,Lop}. Namely, we propose that the lepton doublets $l_2$
and $l_3$ of the second and third families have the same U(1) charges
$A$ and the lepton doublet $l_1$ of the first family has the U(1) charge
$A+1$\cite{SY}, while the right-handed charged leptons $\bar e_i
(i=1-3)$ have the U(1) charges, 2,1,0, respectively(model I). This
lopsided charge assignment for $l_i$ is a crucial point to produce the
large mixing between $\nu_\mu$ and $\nu_\tau$. However, the above choice
of U(1) charge is not unique to reproduce the large $\nu_\mu$ and
$\nu_\tau$ mixing. That is, there is another charge assignment that all
lepton doublets $l_i$ have the same charge $A$(model II)\cite{HKY},
where the right-handed charged leptons $\bar e_i$ have the FN U(1)
charges 3,1,0.

In this letter we show that the above two charge choices lead to
distinct low-energy predictions and hence the two models can be
testable in future neutrino experiments on such as CP violation
in neutrino oscillation and 2$\beta$0$\nu$ decay. We show the FN U(1)
charges $Q_{l_i}$ for $l_i $.
% and $\bar e_i$ i.e $(Q_{e_i})$
 in Table 1. We  assume that the FN U(1) charge for the
Higgs doublet $H$ is zero.

\begin{table}[hbt]
\begin{center}
\begin{tabular}{c|c|c}
&I&II\\
\hline
$Q_{l_3}$ &$A$&$A$\\
$Q_{l_2}$ &$A$&$A$\\
$Q_{l_1}$ &$A+1$&$A$\\
\end{tabular}
 \end{center}
\caption{U(1) charges for lepton doublets}
\label{uptable}
\end{table}

In the FN mechanism the U(1) symmetry is explicitly broken by the vacuum
expectation value of $\phi$, $<\phi>$, where the U(1) charge for $\phi$
is -1. Then all Yukawa couplings are given by the following form:

\begin{eqnarray}
 L&=&h_{ij} \psi_i \psi_j H (\ {\rm or}\ H^*)\left(
\frac{<\phi>}{M_*}\right)^{Q_{\psi_i}+Q_{\psi_j}},
\nonumber\\
&\equiv&
h_{ij} \psi_i \psi_j H (\ {\rm or}\ H^*)\epsilon^{Q_{\psi_i}+Q_{\psi_j}}
\end{eqnarray}
where $h_{ij}$ is a constant with its norm of O(1), $Q_{\psi_i}$ is the FN
 charge for the filed $\psi_i$, $M_*$ the gravitational scale
 $M_*\simeq 2.4\times 10^{18}$ GeV and
 $\epsilon\equiv\frac{<\phi>}{M_*}$.

Then, the neutrino Dirac mass matrix in the model I is given
by\footnote{%
The Dirac mass matrix $M_D$ is defined as
\begin{eqnarray}
L&=&(\nu_L)_i M_{Dij}\bar\nu_R + \ {\rm h.c}.
\nonumber
\end{eqnarray}
}
\begin{eqnarray}
 M_D&=& \epsilon^A m_0
    \left( 
        \begin{array}{ c c c}
         h_{11}\epsilon^{c+1} &h_{12}\epsilon^{b+1} & h_{13}\epsilon^{a+1} \\
         h_{21}\epsilon^{c} &h_{22}\epsilon^{b} & h_{23}\epsilon^{a} \\
         h_{31}\epsilon^{c} &h_{32}\epsilon^{b} & h_{33}\epsilon^{a} 
        \end{array}
    \right)
\label{mdI}\\
&=&\epsilon^A m_0
    \left( 
        \begin{array}{ c c c}
         h_{11}\epsilon &h_{12}\epsilon & h_{13}\epsilon \\
         h_{21} &h_{22} & h_{23} \\
         h_{31} &h_{32} & h_{33} 
        \end{array}
    \right)
    \left( 
        \begin{array}{ c c c}
         \epsilon^{c} &0 & 0 \\
         0 &\epsilon^{b} & 0 \\
         0 &0 &\epsilon^{a} 
        \end{array}
    \right)
\nonumber
\end{eqnarray}
and that in the model II takes the following form
\begin{eqnarray}
 M_D&=& \epsilon^A m_0
    \left( 
        \begin{array}{ c c c}
         h_{11}\epsilon^{c} &h_{12}\epsilon^{b} & h_{13}\epsilon^{a} \\
         h_{21}\epsilon^{c} &h_{22}\epsilon^{b} & h_{23}\epsilon^{a} \\
         h_{31}\epsilon^{c} &h_{32}\epsilon^{b} & h_{33}\epsilon^{a} 
        \end{array}
    \right)
\label{mdII}\\
&=&\epsilon^A m_0
    \left( 
        \begin{array}{ c c c}
         h_{11} &h_{12} & h_{13} \\
         h_{21} &h_{22} & h_{23} \\
         h_{31} &h_{32} & h_{33} 
        \end{array}
    \right)
    \left( 
        \begin{array}{ c c c}
         \epsilon^{c} &0 & 0 \\
         0 &\epsilon^{b} & 0 \\
         0 &0 &\epsilon^{a} 
        \end{array}
    \right)
\nonumber
\end{eqnarray}
The Majorana mass term for right-handed neutrinos $\nu_R$
takes a similar form
\begin{eqnarray}
 M_{\nu_R}&=& M_R
    \left( 
        \begin{array}{ c c c}
         m_{11}\epsilon^{2c} &m_{12}\epsilon^{c+b} & m_{13}\epsilon^{c+a} \\
         m_{12}\epsilon^{b+c} &m_{22}\epsilon^{2b} & m_{23}\epsilon^{b+a} \\
         m_{13}\epsilon^{a+c} &m_{23}\epsilon^{a+b} & m_{33}\epsilon^{2a} 
        \end{array}
    \right)
\label{mR}\\
&=&M_R
    \left( 
        \begin{array}{ c c c}
         \epsilon^{c} &0 & 0 \\
         0 &\epsilon^{b} & 0 \\
         0 &0 &\epsilon^{a} 
        \end{array}
    \right)
    \left( 
        \begin{array}{ c c c}
         m_{11} &m_{12} & m_{13} \\
         m_{12} &m_{22} & m_{23} \\
         m_{13} &m_{23} & m_{33} 
        \end{array}
    \right)
    \left( 
        \begin{array}{ c c c}
         \epsilon^{c} &0 & 0 \\
         0 &\epsilon^{b} & 0 \\
         0 &0 &\epsilon^{a} 
        \end{array}
    \right),
\nonumber
\end{eqnarray}
where we assume that $Q_{\nu_R}$ for right-handed neutrinos are $a,b,c$,
and $m_0$ and $M_R$ represent weak scale and right-handed neutrino mass
scale, respectively. Here $m_{ij}$ is a constant with its norm of O(1)
like the coupling $h_{ij}$.  For numerical convenience we take a basis
where the charged leptons are diagonalized, throughout this letter
except when we determine the $\epsilon$ parameter using the charged
lepton mass matrix (\ref{mlform}).

%For the prediction on CP violation in neutrino
%oscillation and 2$\beta$0$\nu$ decay, the relevant quantity
%is the Majorana mass term for left-handed neutrinos.
From eq.(\ref{mdI}), (\ref{mdII}) and (\ref{mR}) we have the following
Majorana mass term for left-handed neutrinos $\nu_L$\cite{seesaw} in the
model I
\begin{eqnarray}
 &&M_{\nu_L}\label{mlI}\\
&=& \frac{\epsilon^{2A}m_0^2}{M_R}
    \left( 
        \begin{array}{ c c c}
         h_{11}\epsilon &h_{12}\epsilon & h_{13}\epsilon \\
         h_{21} &h_{22} & h_{23} \\
         h_{31} &h_{32} & h_{33} 
        \end{array}
    \right)
    \left( 
        \begin{array}{ c c c}
         m_{11} &m_{12} & m_{13} \\
         m_{12} &m_{22} & m_{23} \\
         m_{13} &m_{23} & m_{33} 
        \end{array}
    \right)^{-1}
    \left( 
        \begin{array}{ c c c}
         h_{11}\epsilon &h_{12}\epsilon & h_{13}\epsilon \\
         h_{21} &h_{22} & h_{23} \\
         h_{31} &h_{32} & h_{33} 
        \end{array}
    \right)^T
\nonumber\\
&\sim& 
   \left( 
        \begin{array}{ c c c}
            \epsilon^2 & \epsilon & \epsilon \\
            \epsilon   & 1        & 1 \\
            \epsilon   & 1        & 1 
        \end{array}
    \right)
\label{symmetry}
\end{eqnarray}
and in the model II
\begin{eqnarray}
 &&M_{\nu_L}\label{mlII}\\
&=& \frac{\epsilon^{2A}m_0^2}{M_R}
    \left( 
        \begin{array}{ c c c}
         h_{11} &h_{12} & h_{13} \\
         h_{21} &h_{22} & h_{23} \\
         h_{31} &h_{32} & h_{33} 
        \end{array}
    \right)
    \left( 
        \begin{array}{ c c c}
         m_{11} &m_{12} & m_{13} \\
         m_{12} &m_{22} & m_{23} \\
         m_{13} &m_{23} & m_{33} 
        \end{array}
    \right)^{-1}
    \left( 
        \begin{array}{ c c c}
         h_{11} &h_{12} & h_{13} \\
         h_{21} &h_{22} & h_{23} \\
         h_{31} &h_{32} & h_{33} 
        \end{array}
    \right)^T
\nonumber\\
&\sim& 
   \left( 
        \begin{array}{ c c c}
            1 & 1 & 1 \\
            1 & 1 & 1 \\
            1 & 1 & 1 
        \end{array}
    \right)
\end{eqnarray}
Note that the FN charges of $\nu_R$'s are irrelevant to the above
$M_{\nu_L}$'s. Therefore, we take $a=b=c=0$ in the present analysis.

We randomly generate the coefficients $h_{ij}$ and $m_{ij}$ such that
their magnitudes be between 0.8 and 1.2\footnote{ In this letter we take
the range between 0.8 and 1.2 for the norm $|h_{ij}|$ and $|m_{ij}|$.
However, the results do not change much even if one takes
a wider range of the norms, say 0.5-1.5.}  and their complex phases be
distributed from 0 to 2$\pi$, and calculate the lepton mixing angles and
the mass square differences for each generated parameters.  We require
that those mixings and the mass square differences satisfy the
conservative constraints from the current experiments\cite{Exp}:

\begin{enumerate}
\item[A.]
$|U_{e3}|< 0.15$ to satisfy the CHOOZ limit.\cite{CHOOZ}
\item[B.]
$4|U_{\mu 3}|^2(1-|U_{\mu 3}|^2)>0.5$
to have the large mixing for atmospheric neutrino oscillation.\cite{SK}
\item[C.]
To satisfy the constraint from solar neutrino deficit,
one of the following two conditions is required to be
     satisfied:\cite{solarK}
\begin{enumerate}
 \item For the small angle solution,
 $10^{-4}<r<10^{-2}$ and
$10^{-4}<\tan^2\theta<5\times 10^{-3}$.
\item for the large angle solution,
$r<0.1$  and  $10^{-1}<\tan^2\theta< 10$.
\end{enumerate}
Here $\tan^2\theta\equiv|U_{e2}/U_{e1}|^2$ and
$r$ is the ratio between the smallest mass square difference and
the second smallest one,
i.e $r\equiv\delta m^2_{\rm solar}/\delta m^2_{\rm atm}$.
\end{enumerate}
Notice that the criterion A is automatically satisfied in the model I,
while $|U_{e3}| \sim $ O(1) generally in the model II.

\begin{figure}[h]
\unitlength 1cm
\begin{picture}(14,10.5)(0,0)
\put(14.1,0.1){\Large $\displaystyle  \epsilon$}
  \includegraphics[width=14cm,height=10.5cm,clip]{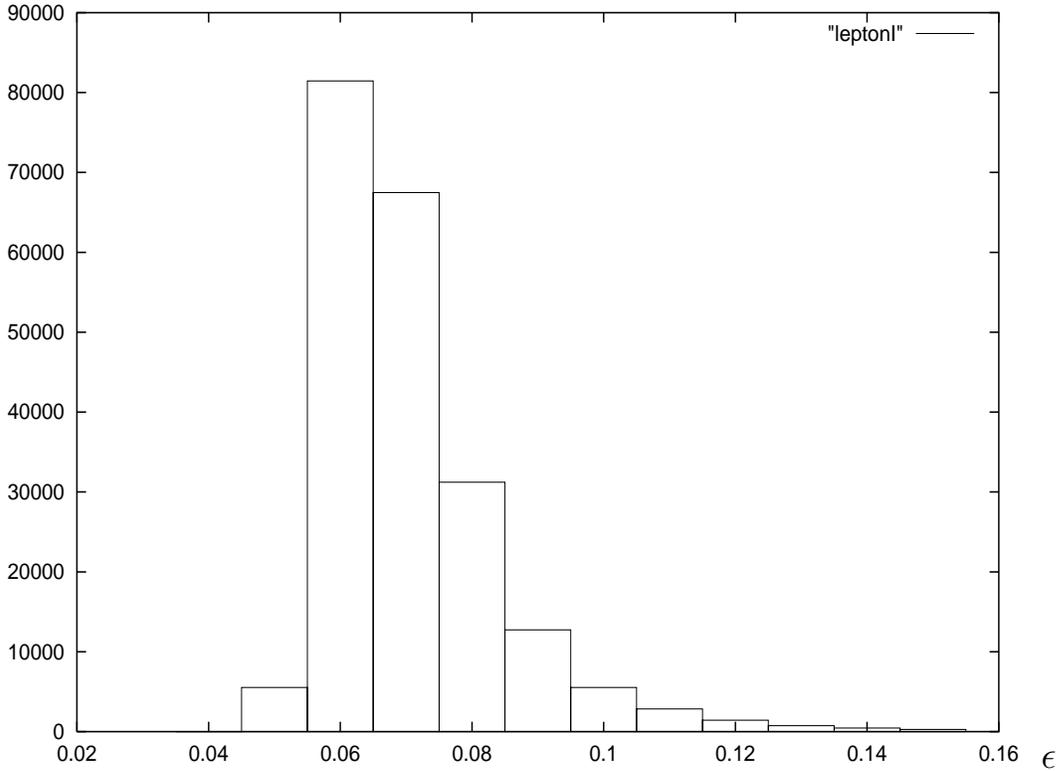}
\end{picture}
\caption{Dependence on $\epsilon$ of how easily we can get 
 solutions for lepton masses. The shape of the dependence does not change
with the tightness of the selection.}
\label{leptoneps}
\end{figure}

To calculate the Majorana masses for left-handed neutrinos
in the model I, we need to fix the value of $\epsilon$.
To find how small value we should take for  $\epsilon$,
we calculate the charged lepton masses with $Q_{e_i}=(0,1,2)$,
\begin{equation}
 M_l \propto 
    \left( 
        \begin{array}{ c c c}
         l_{11}\epsilon^{3} &l_{12}\epsilon^{2} & l_{13}\epsilon^{1} \\
         l_{21}\epsilon^{2} &l_{22}\epsilon^{1} & l_{23}\epsilon^{0} \\
         l_{31}\epsilon^{2} &l_{32}\epsilon^{1} & l_{33}\epsilon^{0} 
        \end{array}
    \right),
\label{mlform}
\end{equation}
where $l_{ij}$'s are randomly generated coefficients
in the same way as $h_{ij}$ and $m_{ij}$.
To see how easily we can have a solution for a given set of coefficients,
we randomly generate 1000000 sets of the coefficients and
find how many sets can satisfy the following cuts.
\begin{eqnarray}
14<&\frac{m_\tau}{m_\mu}&<20
\label{cuts}\\
180<&\frac{m_\mu}{m_e}&<240\nonumber
\end{eqnarray}
The number of sets satisfying these cuts depends on the value of
$\epsilon$.  From the dependence on $\epsilon$ of it in
fig.(\ref{leptoneps}), we can find that $\epsilon$ is likely in the
range between 0.05 and 0.1.\footnote{The absolute values of the vertical
axis depends on how tightly we select samples, so only the shape of the
graph should be considered.} We find that the number of sets in the
model II satisfying the same cuts (eq,(\ref{cuts})) is almost the same
as that in the model I if the right-handed charged leptons $\bar
e_i$ have the FN U(1) charges 3,1,0.

%For simplicity, we use
%$\epsilon = 0.1$ in graphs and show the dependence of the
%predictions on $\epsilon$.

From now on we show the results. We generate one million sets of
coefficients for each $\epsilon=(0.05,0.06,0.07,0.08,0.09,0.1)$ in the
model I and in the model II. Note that the parameter $\epsilon$ is
irrelevant to the neutrino mass matrix (see eq.(\ref{mlII})) in the
model II.  First we see how many sets can remain after the constraints
(A,B and C) are imposed.  It is summarized in table \ref{survive}, where we
list the number of sets separating the cases of small and large mixing
solar neutrino solutions.

\begin{table}[hbt]
\begin{center}
\begin{tabular}{l|c|c}
&small mixing&large mixing\\
\hline\hline
I($\epsilon=0.05$)&81818&20025\\
\hline
I($\epsilon=0.06$)&64353&28837\\
\hline
I($\epsilon=0.07$)&51867&38638\\
\hline
I($\epsilon=0.08$)&42436&49616\\
\hline
I($\epsilon=0.09$)&34714&61220\\
\hline
I($\epsilon=0.1$)&29330&72372\\
\hline\hline
II&6&9703
\end{tabular}
 \end{center}
\caption{Sample of how many sets can satisfy the criterion. Here,
``small(large) mixing'' implies that the set satisfies the criterion for
 the small(large) mixing angle solution to the solar neutrino problem.}
\label{survive}
\end{table}

\begin{figure}[h]
\unitlength 1cm
\begin{picture}(14,9.5)(0,0)
\put(0,9.5){$\displaystyle r=\frac{\delta m^2_{\rm solar}}{\delta m^2_{atm}}$}
\put(13,-0.1){$\displaystyle  |\frac{U_{e2}}{U_{e1}}|^2$}
\put(6,-0.3){model I ($\displaystyle \epsilon=0.1$)}
  \includegraphics[width=13cm,height=9cm,clip]{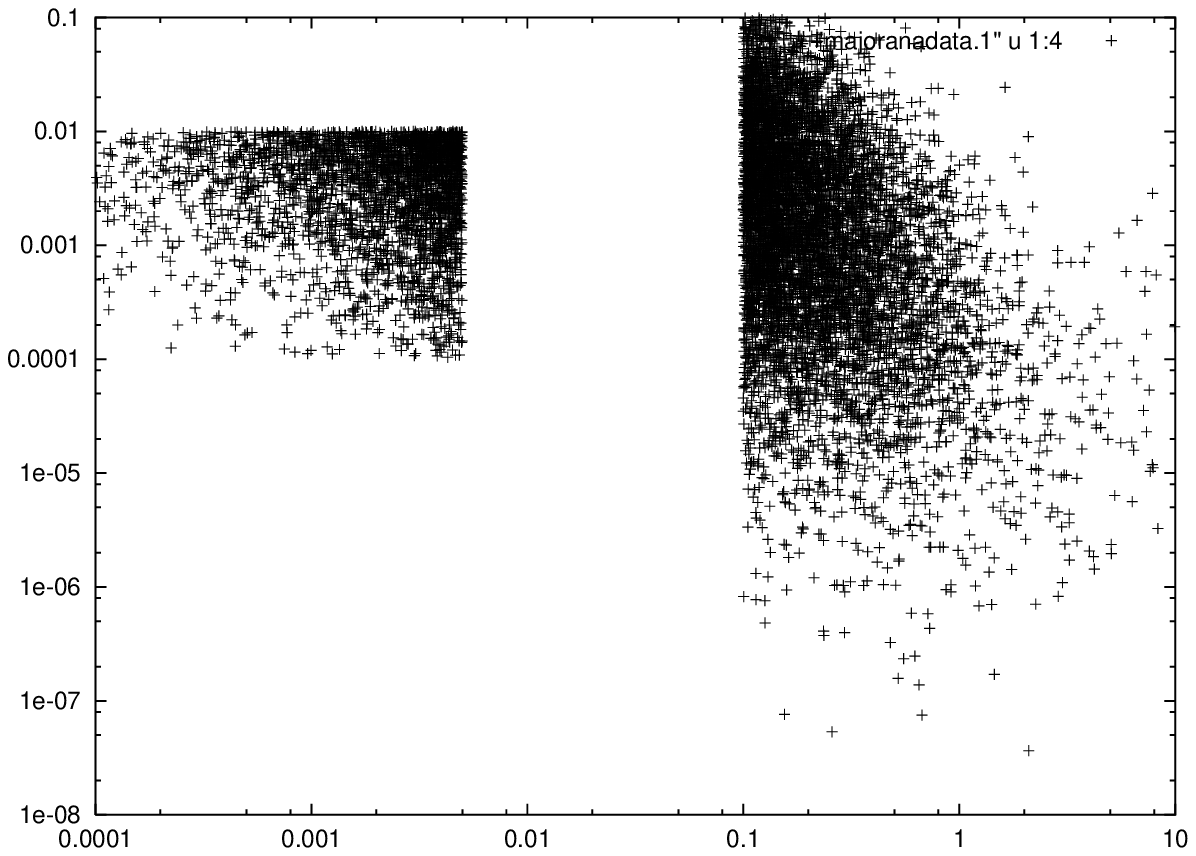}
\end{picture}
\begin{picture}(14,10.5)(0,0)
\put(0,9.5){$\displaystyle r=\frac{\delta m^2_{\rm solar}}{\delta m^2_{atm}}$}
\put(13,-0.1){$\displaystyle  |\frac{U_{e2}}{U_{e1}}|^2$}
\put(6,-0.3){model II }
  \includegraphics[width=13cm,height=9cm,clip]{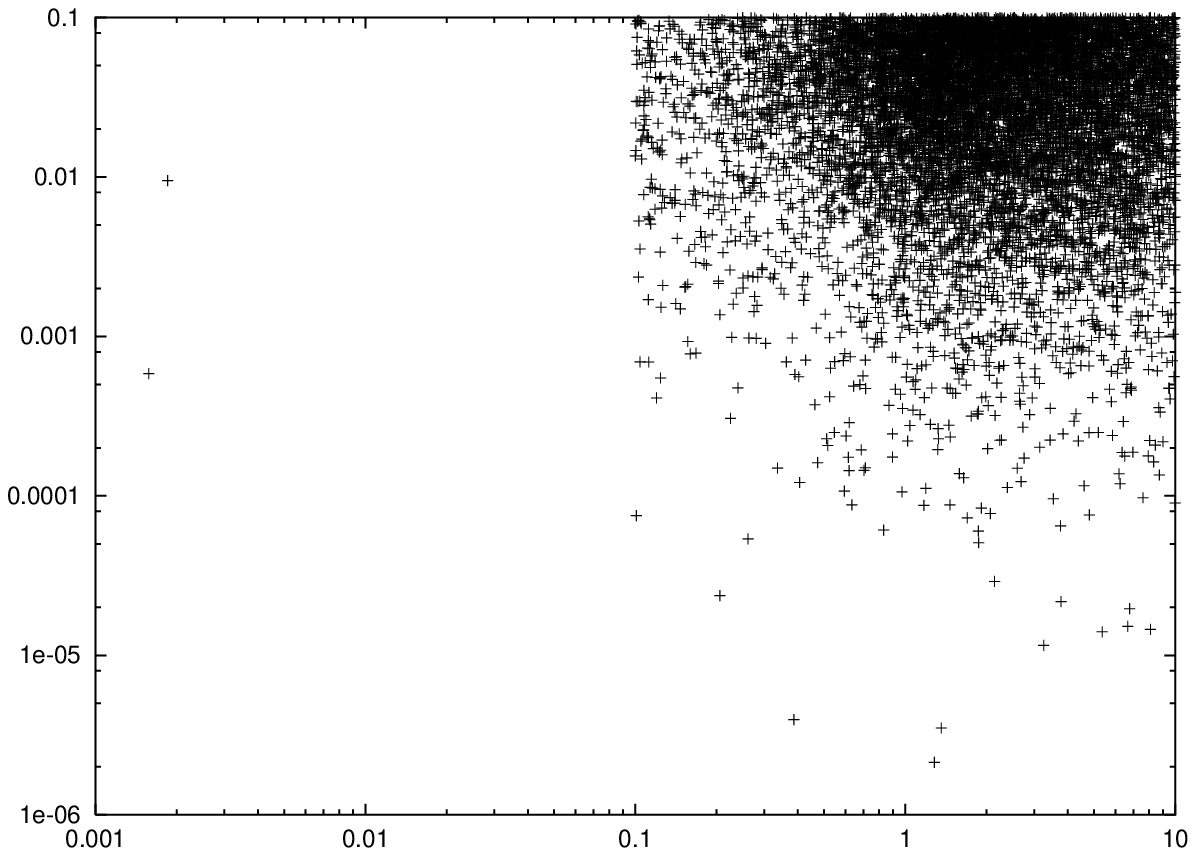}
\end{picture}
\caption{Relations between the mixing angles and mass ratio. The shape of
the dependence does not change much with the tightness of the selection.}
\label{tan2-dm}
\end{figure}

In the model I, we have about 10\% sets of coefficients as solutions
independently of $\epsilon$. This was first pointed out in
ref.\cite{vissani}. On the contrary, in the model II about 1 \% sets can
satisfy the criterion.\footnote{ Note that the model II is essentially
different from the idea of anarchy\cite{anarchy}, where the norms of all
Yukawa couplings vary from 0 to 1. Since they can take very small values
$\sim 0$, the probability realizing small $U_{e3}$ increases.} This
lower probability (1\%) comes mainly from the constraint
$|U_{e3}|<0.15$.  We have also small mixing angle solutions in the model
I, whose physical reason will be discussed later.  In fig. \ref{tan2-dm}
we show the relation between the solar angle $\tan^2\theta\equiv
|U_{e2}/U_{e1}|^2$ and the ratio of the mass square differences $r$.

In the model II $r$ and $\tan^2\theta$ distribute almost uniformly
while in the model I, there is a disfavored region at right top end for
the large angle solutions. This disfavored region can be understood in
the following way. First we note that the Majorana mass matrix takes the
form of eq.(\ref{symmetry}). By diagonalizing the dominant 2 by 2 part,
we will have the following mass matrix,
\begin{equation}
    \left( 
        \begin{array}{ c c c}
            \epsilon^2 & \epsilon & \epsilon \\
            \epsilon   & \delta        & 0 \\
            \epsilon   & 0        & 1 
        \end{array}
    \right),
\label{eq11}
\end{equation}
%The requirement that $r$ should be less than 0.1 in large angle region
%implies $\delta$ is less than 0.3 since roughly 
in which $\delta^2$ corresponds  to $r$ approximately. At the same time
\begin{equation}
 \tan\theta \simeq \epsilon/\delta\ {\rm for}\ \epsilon<\delta. 
\end{equation}
 Thus
\begin{equation}
r\times \tan^2\theta \simeq \epsilon^2
\label{dmtan2}
\end{equation}
and hence $\tan^2\theta>1 $ can be hardly obtained for $r\simeq 0.1$ and 
$\epsilon \simeq 0.1$.

The above arguement also shows the reason why we obtain the large mixing
angle solution for solar neutrino. However it is not complete. Due to
the uncertainties in the right-handed neutrino Majorana mass matrix, the
33 element in eq.(\ref{eq11}) can be rather large and hence there is a
possibility to make $r$ sufficiently small keeping $\delta\sim$ O(1). In
this case the small mixing angle solution may be obtained. This
possibility was not found in \cite{vissani}, since the author of
ref\cite{vissani} considered only the effective operator,
\begin{eqnarray}
L&=&\frac{H^2}{M_R}\kappa_{ij}\nu_{Li}\nu_{Lj}.
\end{eqnarray}

\begin{figure}[h]
\unitlength 1cm
\begin{picture}(14,9.5)(0,0)
\put(13,-0.1){$\displaystyle  |U_{e3}|$}
\put(6,-0.3){model I ($\displaystyle \epsilon=0.1$)}
  \includegraphics[width=13cm,height=9cm,clip]{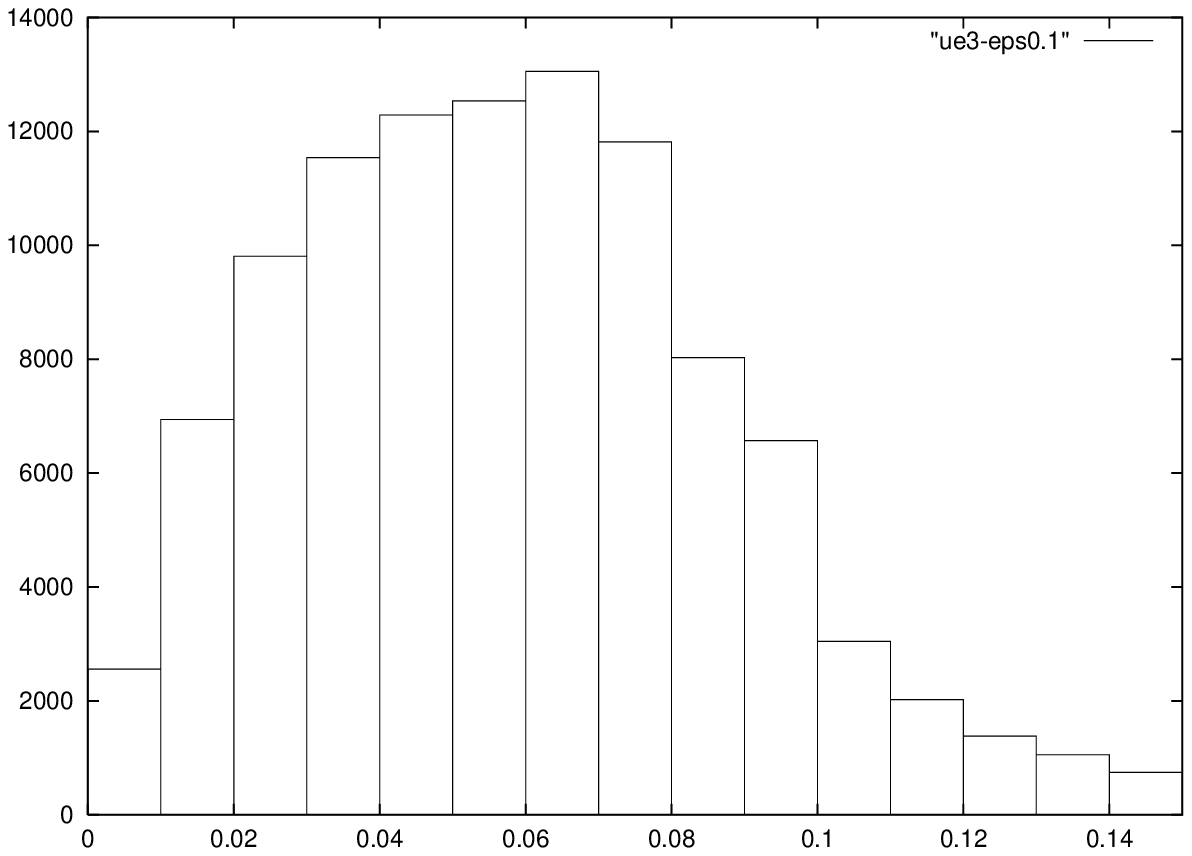}
\end{picture}
\begin{picture}(14,10.5)(0,0)
\put(13,-0.1){$\displaystyle  |U_{e3}|$}
\put(6,-0.3){model II }
  \includegraphics[width=13cm,height=9cm,clip]{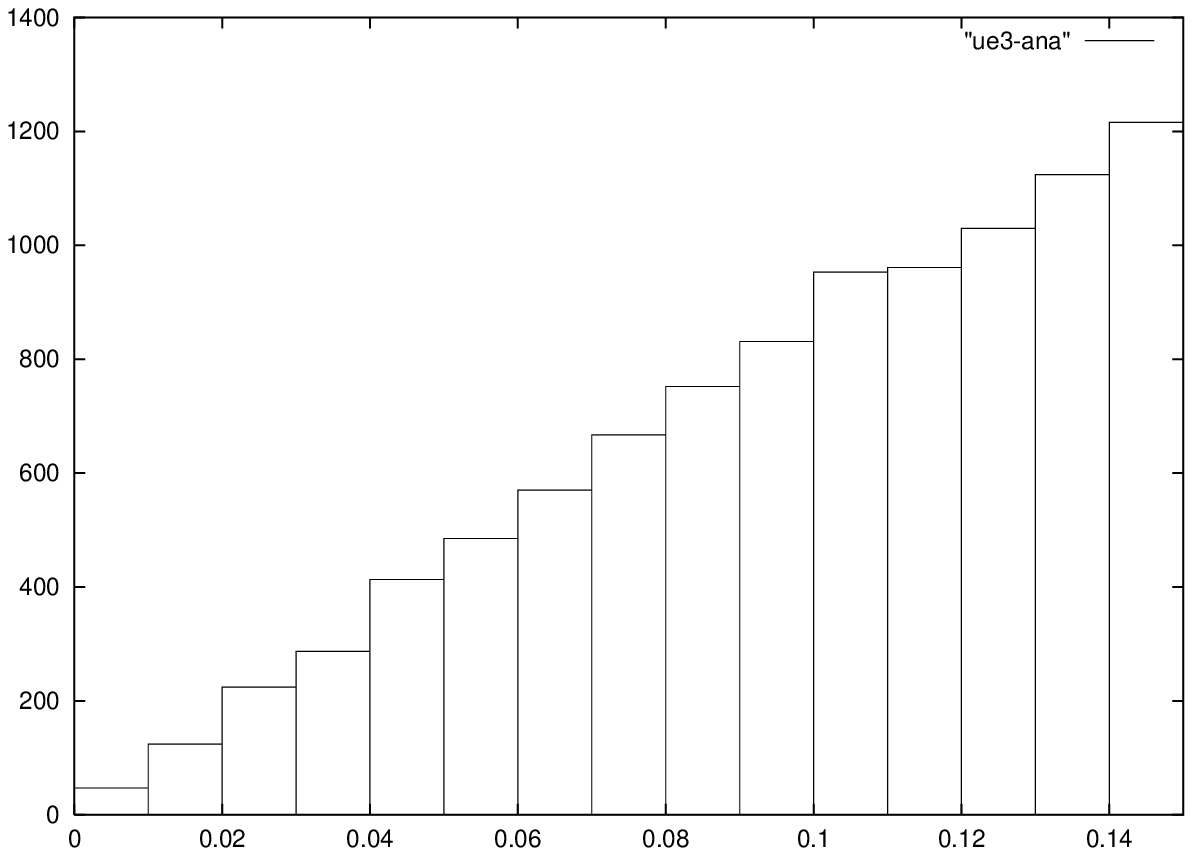}
\end{picture}
\caption{Distributions of $|U_{e3}|$.}
\label{ue3}
\end{figure}

Next we see the distribution of $U_{e3}$ which is one of the most
important parameters in the next generation neutrino oscillation
experiments. In fig. \ref{ue3} we show the distributions of $U_{e3}$ for
both models which satisfy our criterion (A,B and C).  From
eq.(\ref{symmetry}), $U_{e3}$ in the model I is expected to be
proportional to $\epsilon$.  Indeed this scaling can be seen
numerically. On the contrary, since there is no symmetry which
distinguishes the generation in the model II, $U_{e3}$ is likely to be
large and indeed this is seen in fig. \ref{ue3}.  In both models
$U_{e3}$ is expected to be large enough to be observed in future
oscillation experiments like neutrino factory\cite{geer}.

\begin{figure}[h]
\unitlength 1cm
\begin{picture}(13,9.5)(0,0)
\put(13,-0.1){$\displaystyle  J$}
\put(6,-0.3){model I ($\displaystyle \epsilon=0.1$)}
  \includegraphics[width=13cm,height=9cm,clip]{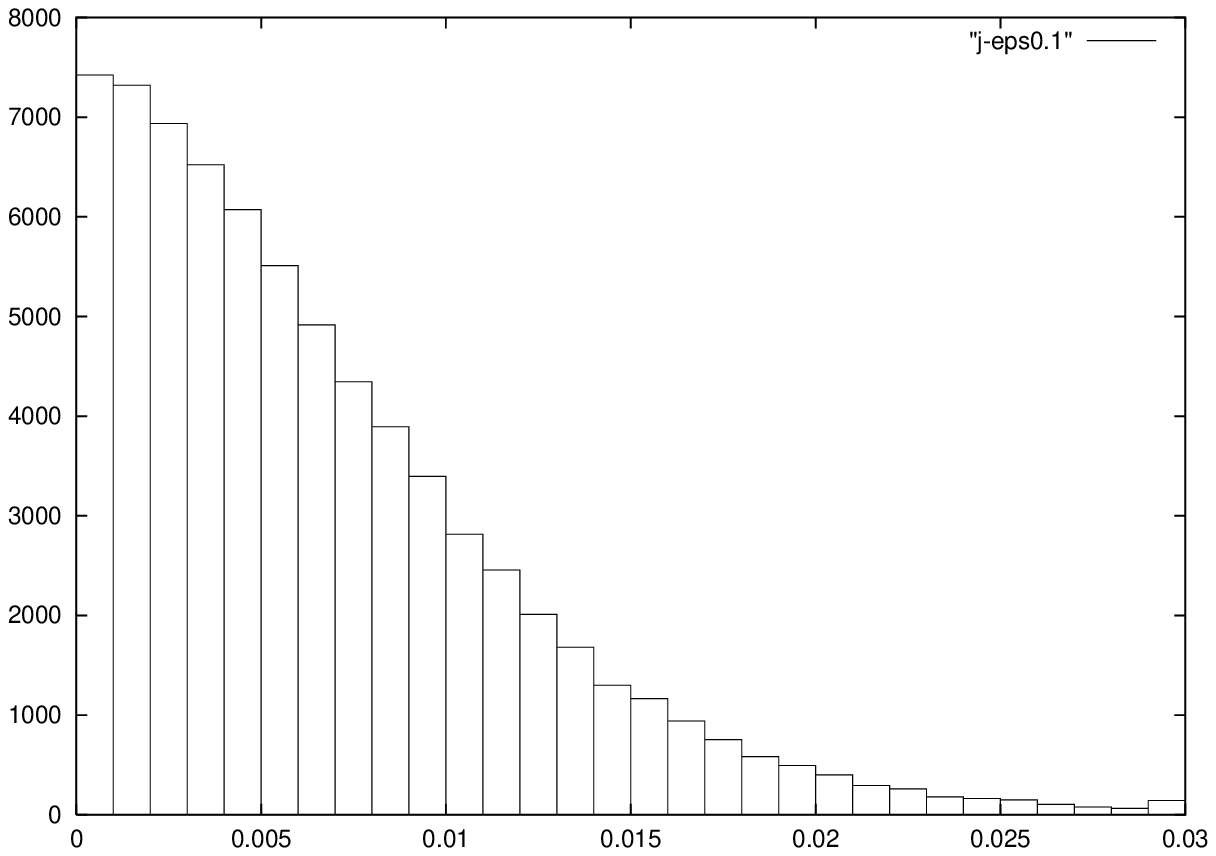}
\end{picture}
\begin{picture}(14,10.5)(0,0)
\put(13,-0.1){$\displaystyle  J$}
\put(6,-0.3){model II }
  \includegraphics[width=13cm,height=9cm,clip]{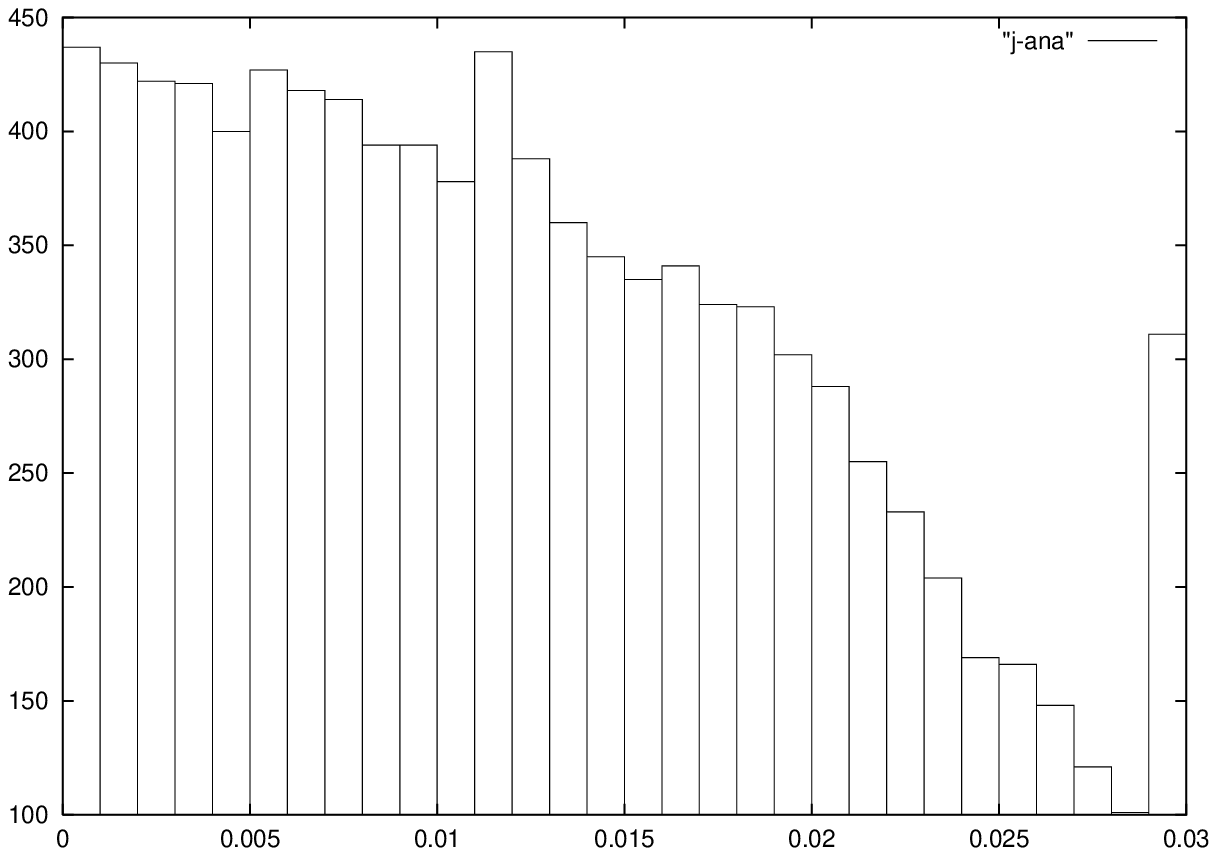}
\end{picture}
\caption{Distributions of $J$. The number of $J$ in the right bin means
that of $J>0.029$.}
\label{J}
\end{figure}

Next we consider how large CP violation can be seen in neutrino
oscillation experiments. The magnitude of CP violation
is characterized by the Jarlskog parameter\cite{jarlskog}
with the Paticle Data Group notation for mixing matrix\cite{PDG}
\begin{eqnarray}
J&\equiv&|{\rm Im}(U^*_{\alpha i} U^*_{\beta j} U_{\alpha j} U_{\beta i})|
\label{defJ}\\
&=&|{\rm Im}(U^*_{e3} U^*_{\mu 2} U_{e2} U_{\mu 3})|\nonumber\\
&=&\frac{1}{4} |\sin\theta_{13} \cos^2\theta_{13} \sin 2\theta_{23}
 \sin 2\theta_{12} \sin\delta|.\nonumber
\end{eqnarray}
Here $\sin\theta_{13}$ is $U_{e3}$, $\theta_{23}$ and $\theta_{12}$
almost correspond to $\theta_{\rm atm}$ and $\theta_{solar}$,
respectively and $\delta$ is the CP violating phase.
%Note that as a function of $\theta_{13}$ $J$ takes its maximum value
%when $\sin\theta_{13}=U_{e3}=1/\sqrt{3}$. 
The distributions of $|J|$ are shown in fig.\ref{J}.  To draw the
graph for the model I, we use only samples which reproduce the large
angle solution for solar neutrino, since samples with the small angle
solution make $J$ much smaller which may not be observable.

The dependence on $\epsilon$ of $J$ in the model I is $J\sim\epsilon$
since $U_{e3}$ is proportional to $\epsilon$ as we see in fig.\ref{ue3},
while the other angles including CP violating phase are of O(1).  That
is,
\begin{eqnarray}
 J &\sim& 
  \begin{array}{ccccc}
0.25&\times&U_{e3}&\times&{\rm\ other\ contribution\ in
\ eq.(\ref{defJ}) } \\
{\rm prefactor}& &\epsilon&&{\rm O(1)}\\
  \end{array}\\
&\simeq&0.1\epsilon.
\nonumber
\end{eqnarray}
%Here we roughly estimate  the contributions by the terms of O(1)
%reduce $J$ by a several factor.
On the other hand, $J$ in the model II is estimated to be
\begin{eqnarray}
 J &\sim& 
  \begin{array}{ccccc}
0.25&\times&U_{e3}&\times&{\rm\ other\ contribution\ in
\ eq.(\ref{defJ}) } \\
{\rm prefactor}& &0.15&&{\rm O(1)}\\
  \end{array}\\
&\sim& 0.01.
\nonumber
\end{eqnarray}
and hence slightly larger than that in the model I.

In the realistic neutrino oscillation experiment, the measurable
quantity for CP violation is not $J$ itself but\cite{AKS}
\begin{equation}
 \tilde J \equiv |J\times \frac{\delta m^2_{\rm solar}}{\delta m^2_{\rm atm}}|.
\end{equation}

\begin{figure}[h]
\unitlength 1cm
\begin{picture}(14,9.5)(0,0)
\put(13,-0.1){$\displaystyle  \tilde J$}
\put(6,-0.3){model I ($\displaystyle \epsilon=0.1$)}
  \includegraphics[width=13cm,height=9cm,clip]{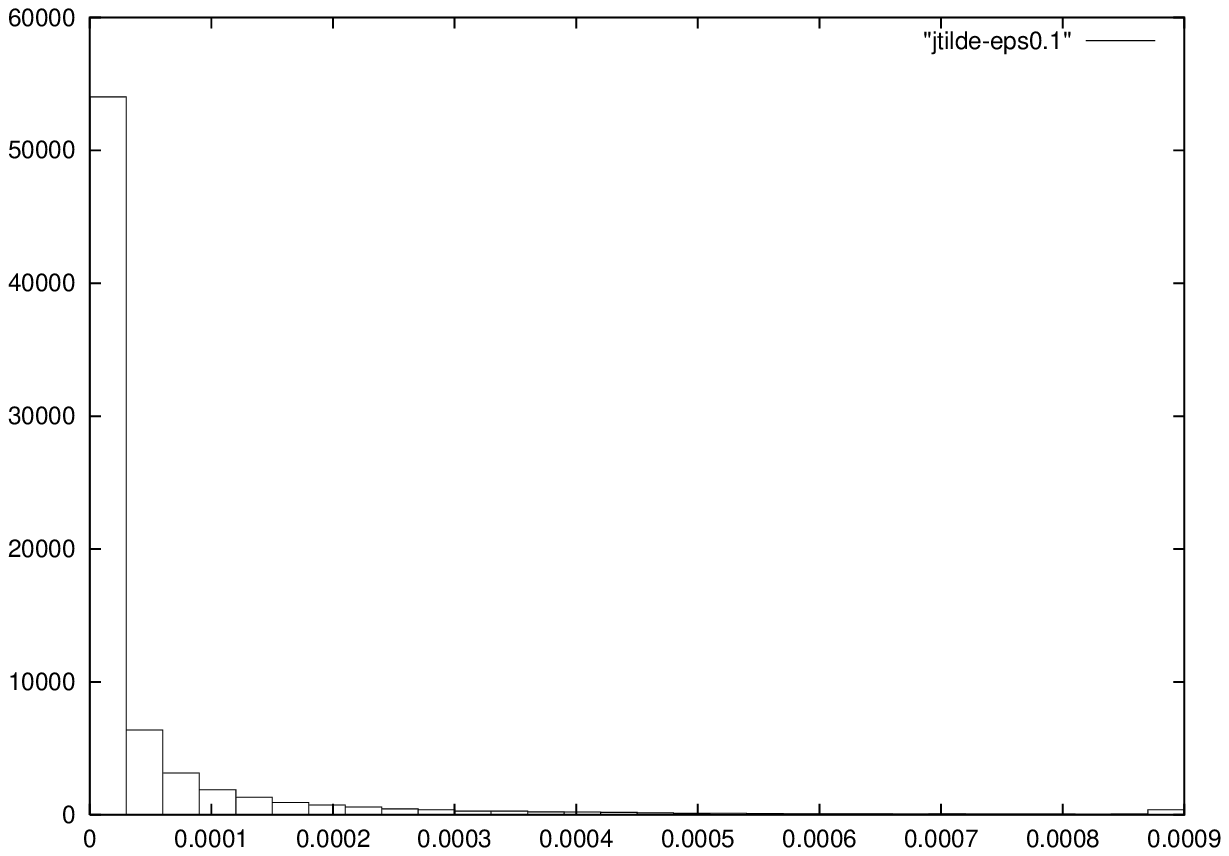}
\end{picture}
\begin{picture}(14,10.5)(0,0)
\put(13,-0.1){$\displaystyle  \tilde J$}
\put(6,-0.3){model II }
  \includegraphics[width=13cm,height=9cm,clip]{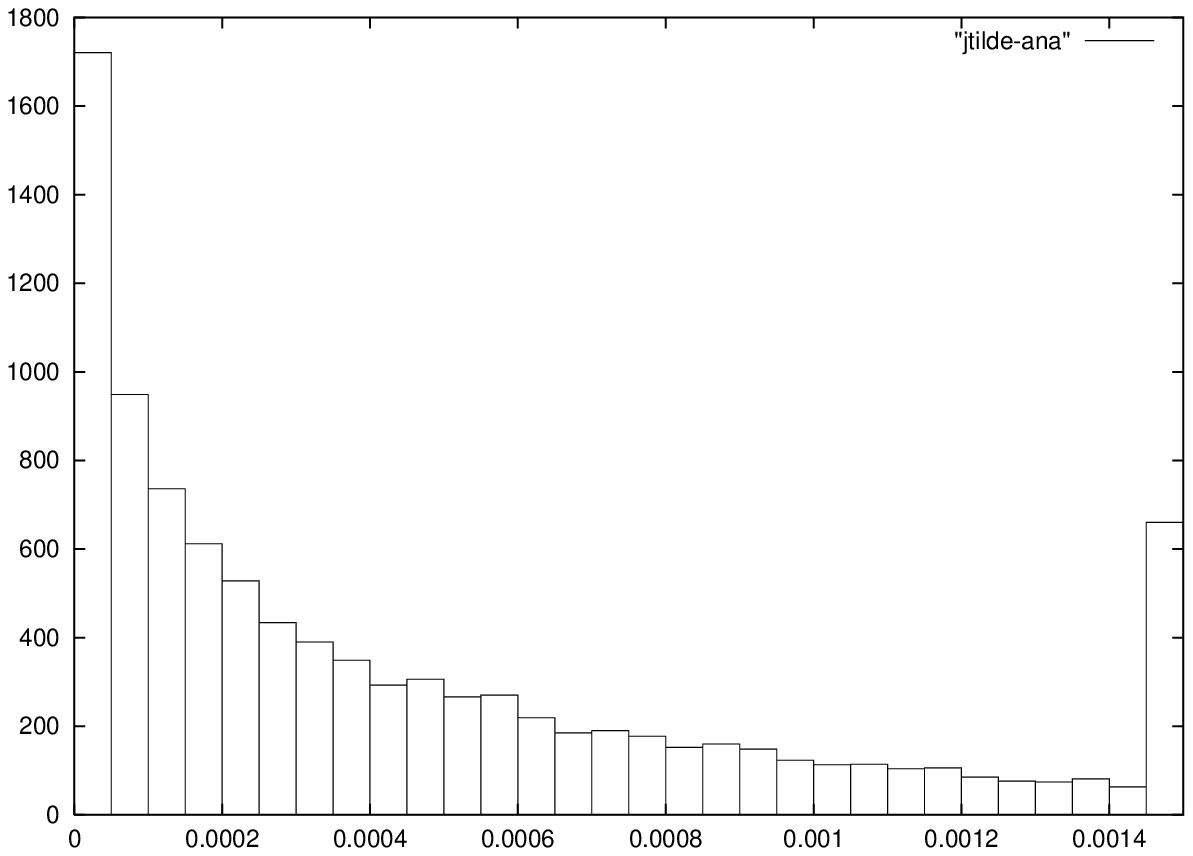}
\end{picture}
\caption{Distributions of $\tilde J \equiv |J\times r|$.
The number of $\tilde J$ in the right bin means
that of $\tilde J>0.0009$ for the model I and $\tilde J>0.00145$ for the 
 model II.}
\label{tildeJ}
\end{figure}

In fig.\ref{tildeJ} we plot the distribution of $|\tilde J|$.
The dependence on $\epsilon$ of $\tilde J$ in the model I is rather
complicated. Due to eq.(\ref{dmtan2}), 
\begin{eqnarray}
\tilde J &\sim& 
  \begin{array}{ccccccc}
0.5&\times&U_{e3}&\times&
\frac{\delta m^2_{\rm solar}}{\delta m^2_{\rm atm}} \tan\theta_{12}
&\times&{\rm\ other\ contribution\ in
\ eq.(\ref{defJ}) } \\
{\rm prefactor}& &\sim\epsilon&&
\epsilon^2/\tan\theta_{12}&&{\rm O(1)}\\
  \end{array}\\
&\simeq&0.1\times \epsilon^3.
\nonumber
\end{eqnarray}
On the contrary $\tilde J$ in the model II
is estimated easily as
\begin{equation}
 J\times\frac{\delta m^2_{\rm solar}}{\delta m^2_{\rm atm}} \sim 0.01 \times
(0.1 - 0.01) \sim ( 0.001 - 0.0001 ).
\end{equation}
Thus, there is a possibility to measure CP violation in the next
generation neutrino oscillation experiments for the model II while there
seems to be little hope to see CP violation in the near future
experiments\cite{CP} for the model I.

\begin{figure}[h]
\unitlength 1cm
\begin{picture}(14,9.5)(0,0)
\put(13,-0.1){$\displaystyle  m_{ee}$ [eV]}
\put(6,-0.3){model I ($\displaystyle \epsilon=0.1$)}
  \includegraphics[width=13cm,height=9cm,clip]{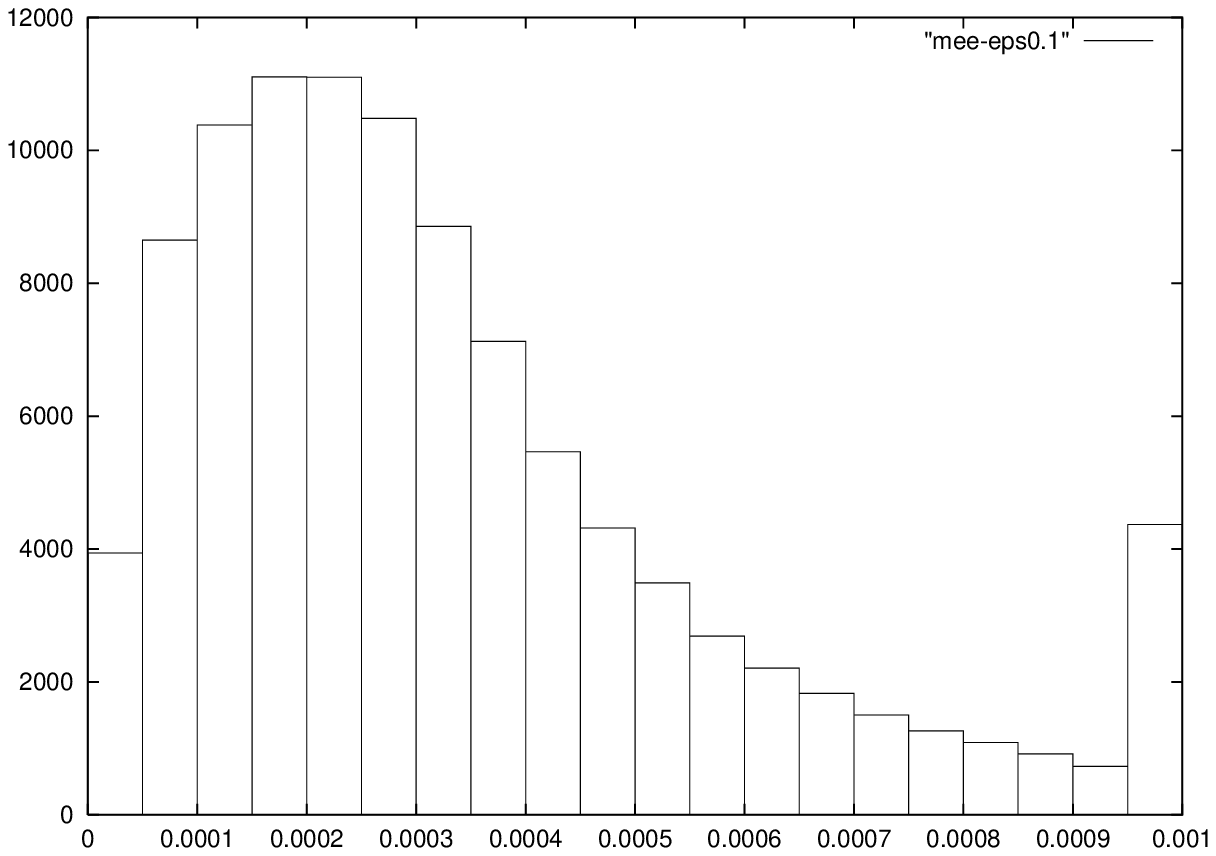}
\end{picture}
\begin{picture}(14,10.5)(0,0)
\put(13,-0.1){$\displaystyle  m_{ee}$ [eV]}
\put(6,-0.3){model II }
  \includegraphics[width=13cm,height=9cm,clip]{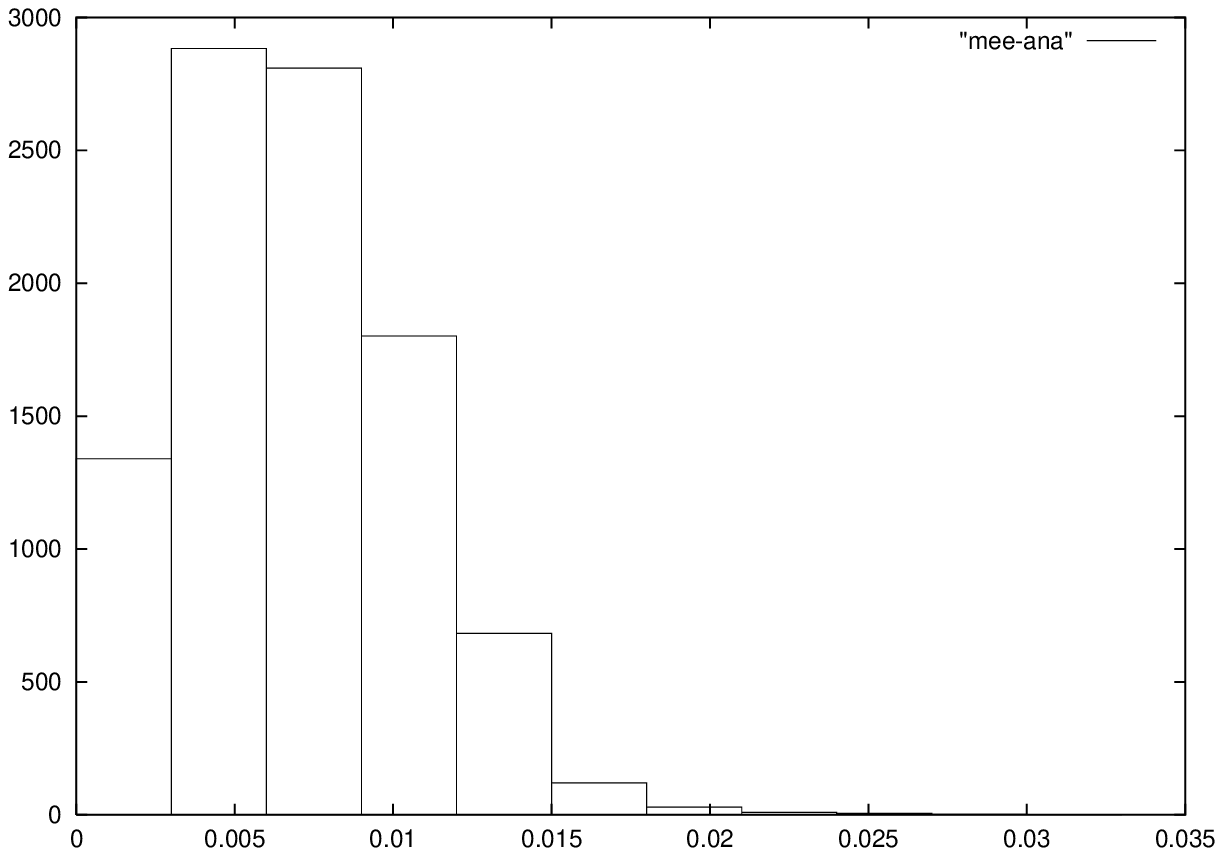}
\end{picture}
\caption{Distributions of $m_{ee}$.The number of $m_{ee}$ in the right
 bin for the model I means that of $m_{ee}>0.00095$ eV.}
\label{mee}
\end{figure}

Finally, we calculate $m_{ee}\equiv |\sum_i U^2_{ei}m_i|$ which is a crucial
parameter to determine the 2$\beta$0$\nu$ decay rate.  In fig.\ref{mee}
we show the distributions of $m_{ee}$ in the both models.  In the
figure, we have assumed $\delta m^2_{\rm atm}=3\times 10^{-3}$ eV$^2$.
The dependence on $\epsilon$ of $m_{ee}$ in the model I is very simple.
Due to eq.(\ref{symmetry}), 
\begin{eqnarray}
m_{ee} &\sim& \sqrt{3\times 10^{-3}} \epsilon^2
\label{meeI}\\
&\simeq&0.05\times \epsilon^2\ {\rm eV}.
\nonumber
\end{eqnarray}
On the contrary $m_{ee}$ in the model II is naively expected to be
$\sqrt{3\times 10^{-3}}\sim 0.05$ eV.  However the samples which satisfy
our criterion prefer values lower by almost one order of magnitude.

Again, there is a possibility to find the 2$\beta$0$\nu$ decay in the
next generation experiments for the model II while there is little hope to
see it in the near future experiments for the model I.\cite{nuless}

In conclusion, we summarize the results. There are two kinds of FN U(1)
charges which realize lopsided structure for the lepton doublets mass
matrices, i.e the model I (001) and the model II(000). These two sets of
charge assignments have very different feature from each other and hence
it is testable in the near future experiments which type is likely the
case.

We have considered first the solar neutrino oscillation.  In the model
II we hardly get a small angle solution to the solar neutrino problem,
so if the solar neutrino deficit is explained by the small angle
solution, then the model II will be rejected. In the model I there is a
disfavored region, which is explained by eq.(\ref{dmtan2}), and hence if
it explains the solar neutrino deficit then the model I will be
disfavored.

Next we have discussed the distribution of $U_{e3}$. Unfortunately
as is seen in fig.\ref{ue3} it is very difficult  to distinguish the
model I from the model II by this angle.

Then we have studied CP violation in the lepton sector. As is shown in
fig's.\ref{J} and \ref{tildeJ}, it seems difficult to see CP violation
in the next generation neutrino oscillation experiments in the model I,
while there is a possibility to observe it in the model II. Therefore,
if we detect the CP violation in the lepton sector then the model II
will be favored.

Finally, we have examined how large $m_{ee}\equiv \sum_i U^2_{ei}m_i$
can be, which is a key element for 2$\beta$0$\nu$ decay. The
distribution of it in the model I shows the expected shape from
eq.(\ref{meeI}) as is seen in fig\ref{mee}, while in the model II its
magnitude is smaller almost by one order of magnitude than that naively
expected.  However, $m_{ee}$ in the model II lies in the range
accessible in the near future experiments while it will be harder to see
2$\beta$0$\nu$ decay in the model I.

%%%%%%%%%%%%%%%%%%%%%%%%%%%%%%%%%%%%%%%%%%%%%%%%%%%%%%%%%%%%%%%%%%%%%%
\subsection*{Acknowledgments}
%%%%%%%%%%%%%%%%%%%%%%%%%%%%%%%%%%%%%%%%%%%%%%%%%%%%%%%%%%%%%%%%%%%%%%

The authors are grateful to N. Haba and H. Murayama for useful discussions.
The work of J. S is supported  in part by a Grant-in-Aid for Scientific
Research of the Ministry of Education, Science and Culture,
\#12047221, \#12740157.
The  work of T.Y is supported in part by the Grant-in-Aid, 
Priority Area ``Supersymmetry and Unified Theory of Elementary
Particles''(\#707).

%%%%%%%%%%


\begin{thebibliography}{99}
\bibitem{FN} C.D.~Froggatt and H.B.~Nielsen, Nucl. Phys. {\bf B147}
(1979) 277.

\bibitem{SK}
Y.~Fukuda {\it et al.} [Superkamiokande Collaboration],
Phys. Lett. {\bf B433} (1998) 9;
Phys. Lett. {\bf B436} (1998) 33;
Phys. Rev. Lett. {\bf 81} (1998) 1562.
    
\bibitem{SY} J. ~Sato and T. ~Yanagida, Phys. Lett. {\bf B430} (1998) 127;
	Nucl. Phys. B (Proc. Suppl.) 77 (1999) 293.\\
    W. Buchmuller and T. Yanagida,  Phys. Lett. {\bf B445} (1999) 399.

\bibitem{Lop}C. H. Albright, K.S. Babu and S.M. Barr ,
Phys. Rev. Lett. {\bf 81} (1998) 1167.\\
N. Irges, S. Lavignac and P. Ramond,
 Phys. Rev. {\bf D58} (1998) 035003.

\bibitem{HKY}
J.~Hisano , K. ~Kurosawa and Y. ~Nomura, 
Nucl. Phys. {\bf B584} (2000) 3.

\bibitem{seesaw}
T. Yanagida, 
{\it in} Proc. Workshop on the unified theory and 
the baryon number in the universe, (Tsukuba, 1979), 
{\it eds.} O. Sawada and S. Sugamoto,
Report KEK-79-18 (1979);\\
M. Gell-Mann, P. Ramond and R. Slansky,
{\it in} ``Supergravity''
(North-Holland, Amsterdam, 1979)
{\it eds.} D.Z. Freedman and P. van Nieuwenhuizen.

\bibitem{Exp} G.L. Fogli, E. Lisi, A. Marrone and G. Scioscia
Phys. Rev. D59 (1999) 033001\\
M.C. Gonzalez-Garcia and C. Pea-Garay, hep-ph/0009041.

\bibitem{CHOOZ} 
M. Apollonio {\it et al.}, Phys. Lett. {\bf B466} (1999) 415.

\bibitem{solarK} Super--Kamiokande Collaboration, Y. Fukuda {\it et al.},
 Phys. Rev. Lett. {\bf 81} (1998) 1158 ; Erratum {\bf 81} (1998) 4279 ; 
 {\bf 82} (1999) 1810; {\bf 82} (1999) 2430 ;
  Y. Suzuki, Nucl. Phys. B (Proc. Suppl.) {\bf 77} (1999) 35.

\bibitem{vissani}F. Vissani, JHEP 9811 (1998) 025.

\bibitem{anarchy}L. Hall, H. Murayama and N. Weiner,
Phys. Rev. Lett. 84 (2000) 2572.

\bibitem{geer} S. Geer, Phys. Rev. {\bf D57} (1998) 6989,
Erratum-ibid {\bf D59} (1999) 039903.

\bibitem{jarlskog}  C. Jarlskog,  Phys. Rev. Lett. {\bf 55} (1985) 1039.

\bibitem{PDG}
Particle Data Group Eur. Phys. J. C15 (2000) 1.

\bibitem{AKS}
J.~Arafune, M.~Koike and J.~Sato,
  Phys. Rev. {\bf D56 } (1997) 3093, Erratum-ibid. {\bf D60} (1999) 119905.

\bibitem{CP}J Sato,  hep-ph/0008056\\
 B. Richter, hep-ph/0008222 \\
A. Cervera, A. Donini, M.B. Gavela, J.J. Gomez Cadenas, P. Hernandez,
O. Mena and S. Rigolin,  Nucl. Phys. {\bf B579} (2000)17.

\bibitem{nuless} J.~Hellmig and H. V. Klapdor-Kleingrothaus,
Z. Phys. {\bf A359} (1997) 351.\\
       Klapdor-Kleingrothaus and M. Hirsch,
Z. Phys. {\bf A359} (1997) 361.
 
\end{thebibliography}
\end{document}